\documentclass[a4paper,12pt]{article}

\usepackage[english]{babel}
\usepackage{color}
\usepackage{amssymb}
\usepackage{amsmath}
\usepackage{graphicx}
\usepackage[small,bf]{caption}
\usepackage{geometry}
\usepackage{hyperref}
\usepackage[T1]{fontenc}

\usepackage{subcaption}

\begin{document}

\vspace*{-15mm}
\begin{flushright}
TTP24-012\\
P3H-24-029
\end{flushright}
\vspace*{1cm}

\begin{center}  
\textbf{{\huge Towards understanding fermion masses and mixings}}

\vspace{1cm}

Zurab Berezhiani$^{a,b,}\footnote{zurab.berezhiani@aquila.infn.it}$, Benedetta Belfatto$^{c,}\footnote{benedetta.belfatto@kit.edu}$  \\[6mm]
{\it $^a$ Dipartimento di Scienze Fisiche e Chimiche, Universit\`a di L'Aquila,\\ via Vetoio 1, 
Coppito, 67100 L'Aquila}\\[1mm]
{\it $^b$ INFN, Laboratori nazionali del Gran sasso, Assergi, 67100 L'Aquila, Italy}\\[1mm]
{\it $^c$ Institut f\"ur Theoretische Teilchenphysik, Karlsruhe Institute of Technology,\\ Engesserstra{\ss}e 7, D-76128 Karlsruhe, Germany}\\[1mm]

\end{center}

\begin{abstract}
The Standard Model does not constrain the form of the Yukawa matrices and thus
the origin of fermion mass hierarchies and mixing pattern remains puzzling.
On the other hand, there are intriguing relations between the quark 
masses and their weak mixing angles, such as the well-known one  
$\tan \theta_C= \sqrt{m_d/m_s}$ for the Cabibbo angle,   
which may point towards specific textures of Yukawa matrices hypothesized
by Harald Fritzsch at the end of the 70's. 
Though the original ansatz of Fritzsch is excluded by the experimental data,  
one can consider its minimal modification which consists 
in introducing an asymmetry between the 23 and 32 entries 
in the down-quark Yukawa matrix.  
We show that this structure is perfectly compatible with the present precision data
on quark masses and CKM mixing matrix, and  theoretically it 
can be obtained  in the context of $SU(5)$ model 
with inter-family $SU(3)_H$ symmetry. 
We also discuss some alternative approaches which could give a natural 
description of the fermion mass spectrum and weak mixing pattern. 
\end{abstract}

 \newpage


\section{Introduction}

The replication of fermion families is one of the main puzzles of particle physics. 
Three fermion families are in identical representations of the Standard Model (SM) 
gauge symmetry $SU(3)\times SU(2)\times U(1)$.
The left-handed (LH) quarks $q_{Li}=(u_L,d_L)_i$ and leptons $\ell_{Li}=(\nu_L,e_L)_i$  
transform as weak doublets while right-handed (RH) ones  
$u_{Ri},d_{Ri},e_{Ri}$ are weak singlets, $i=1,2,3$ being the family index. 
The SM contains the unique order parameter -- vacuum expectation value (VEV)
of the Higgs doublet  $\phi$, $\langle \phi^0 \rangle = v_{\rm w} = 174$ GeV, 
which spontaneously breaks the electroweak symmetry $SU(2)\times U(1)$.  
It determines  the mass scale of the  weak bosons $W^\pm,Z$ as well as fermion masses 
which emerge  via the Yukawa couplings 
\begin{align}
\label{Yukawas-SM}
 Y_u^{ij}   u^c_i  q_j  \phi  + Y_d^{ij}  d^c_i q_j \tilde\phi   + Y_e^{ij}  e^c_i \ell_j \tilde\phi
 \, + \, {\rm h.c.}  
\end{align}
where $Y_{e,u,d}$ are the Yukawa coupling matrices, and $\tilde\phi=i\tau_2 \phi^\ast$.  
Here we use instead of the RH fermion fields their complex conjugates
as $u^c_{L}= {\rm C} \overline{u_{R}}^T$ (anti-fields) and omit in the following 
the subscript $L$ for  $q$, $u^c$, $d^c$ etc., all being LH Weyl spinors.\footnote{ 
With these notations, the description can be conveniently extended for a 
supersymmetric version of the SM,  
in which case  the terms \eqref{Yukawas-SM}  correspond to 
the Yukawa superpotential, with $\phi=\phi_u$ and $\tilde\phi = \phi_d$
being two Higgs doublets  taken as left-chiral superfields.}
The Yukawa couplings \eqref{Yukawas-SM},  after substituting the Higgs VEV $v_w$, 
induce the fermion mass matrices $M_f = Y_f v_{\rm w}$ ($f = u,d,e$) which are 
generically non-diagonal.  They can be brought to the diagonal form 
(i.e. to the mass eigenstate basis)  via bi-unitary transformations: 
\begin{align}
& U_{f}^{\dagger} M_f V_{f} = M_f^{\rm diag} 
\label{diag}
\end{align}
so that the quark masses  $m_u,m_c,m_t$ and $m_d,m_s,m_b$   
are the eigenvalues of the mass matrices $M_u$ and $M_d$, 
as the charged lepton masses $m_e,m_\mu,m_\tau$ are the eigenvalues of $M_e$  
(we shall not discuss the neutrino masses in this paper).  

In the SM context, the matrices $U_{u,d}$ which rotate the RH fermions  
have no physical meaning while the ``left" ones $V_{u,d}$ give rise to the mixing
in the quark charged currents coupled to weak $W^\pm$ bosons. 
This mixing  is described by the unitary Cabibbo-Kobayashi-Maskawa (CKM) $V_\text{CKM}$
\cite{Cabibbo:1963yz,Kobayashi:1973fv}:
\begin{align}\label{CKM}
V_\text{CKM}= V_{u}^\dagger V^{\,}_{d}= \left(\begin{array}{ccc}
V_{ud} & V_{us} & V_{ub}  \\
V_{cd} & V_{cs} & V_{cb}  \\
V_{td} & V_{ts} & V_{tb} 
\end{array}\right) 
\end{align}
By rotating away the irrelevant phases,  the unitary matrix $V_\text{CKM}$ 
can be conveniently parameterized in terms of four parameters, 
 three mixing angles and a CP-violating phase  \cite{Kobayashi:1973fv}.
In the standard parameterization adopted by Particle Data Group  (PDG)
\cite{Workman:2022ynf},   
the  angles  are chosen as $\theta_{12}$, $\theta_{23}$,  $\theta_{13}$, 
and the CKM matrix reads 
 \begin{align}
 \label{ckmSP}
V_\text{CKM}=  \left(\begin{array}{ccc}
c_{12}c_{13} & s_{12}c_{13} & s_{13}e^{-i\delta} \\
-s_{12}c_{23}-c_{12}s_{23}s_{13}e^{i\delta} & c_{12}c_{23}-s_{12}s_{23}s_{13}e^{i\delta} 
& s_{23}c_{13}  \\
s_{12}s_{23}-c_{12}c_{23}s_{13}e^{i\delta} & -c_{12}s_{23}-s_{12}c_{23}s_{13}e^{i\delta} 
& c_{23}c_{13}
\end{array}\right) 
\end{align}
where $s_{ij}=\sin\theta_{ij}$, $c_{ij}=\cos\theta_{ij}$ and $\theta_{ij}$ 
are chosen so that $s_{ij},c_{ij}\geq 0$.
As a measure of CP violation, the rephasing-invariant quantity
$J\sum_{m,n}\epsilon_{ikm}\epsilon_{jln}=\text{Im}[V_{ij}V_{kl}V_{il}^*V_{kj}^*]$  
(the Jarlskog invariant) \cite{Jarlskog:1985ht} in the standard parameterization reads:
\begin{align}\label{J-inv}
 J=\sin\delta\, s_{12}s_{23}s_{13} \, c_{12}c_{23}c_{13}^2 
\end{align}

The SM has a remarkable feature: the natural flavor conservation in neutral currents, 
namely in the fermion couplings with the Higgs and $Z$ bosons  
\cite{Glashow:1970gm,Glashow:1976nt,Paschos:1976ay}.  
However,  it contains no theoretical input that could explain  
the fermion mass spectrum and the weak mixing pattern. 
In a sense, the SM is technically natural since it can tolerate any  pattern
of the Yukawa matrices  $Y_{u,d,e}$  but it can tell nothing on the origin 
of the  strong hierarchies between their eigenvalues as well as of the 
 nearly aligned structures  of the matrices $Y_u$ and $Y_d$.  
 This remains true also in the context of supersymmetric and/or grand unification theories (GUT). 
 So, the origin of the fermion mass and mixing pattern remains a mystery.  
 
\begin{figure}[t]
\centering
\includegraphics[width=2.0in]{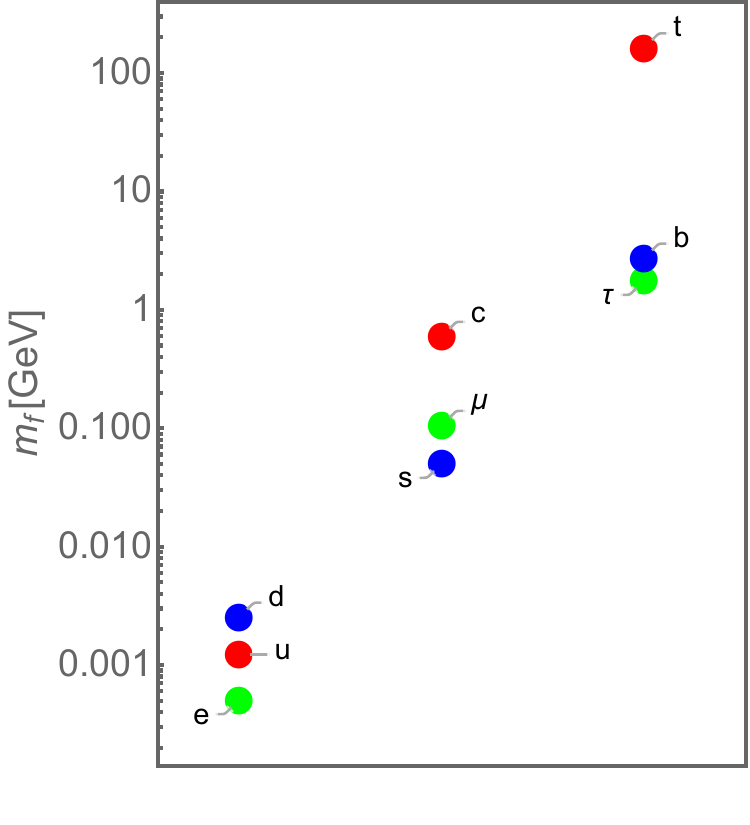}
\caption{ The running masses of quarks and leptons,  
normalized  at the electroweak scale $\mu = v_w$. 
}
\label{plotmasse}
\end{figure}

The quark and lepton mass spectrum, schematically shown in fig. \ref{plotmasse}, 
exhibits a strong inter-family hierarchy. 
The mass hierarchy between the third and first families 
is $m_b/m_d \sim 10^3$ for the down quarks,  
and yet stronger for the up quarks, $m_t/m_u \sim 10^5$. 
 Expressed in terms of the small parameters 
$m_d/m_s=\epsilon_d$,  $m_s/m_b=\epsilon_s$ etc., these hierarchies 
approximately look as:
 \begin{align}\label{mass-ratios} 
& m_t : m_c : m_u = 1 : \epsilon_c : \epsilon_u \epsilon_c \, , \quad 
 \epsilon_c \approx \frac{1}{300},  ~~ \epsilon_u \approx \frac{1}{500}
 ~ \to ~ \epsilon_u  \epsilon_c \approx \frac{1}{150000} 
\nonumber  \\  
& m_b : m_s : m_d = 1 : \epsilon_s : \epsilon_d \epsilon_s\, , \quad 
~ \epsilon_s \approx \frac{1}{50},  ~~~ \epsilon_d \approx \frac{1}{20} 
~~ \to ~ \epsilon_d  \epsilon_s \approx \frac{1}{1000} \\
& m_\tau : m_\mu  : m_e = 1 : \epsilon_\mu : \epsilon_e \epsilon_\mu  \, , \quad 
 \epsilon_\mu \approx \frac{1}{17},  ~~ \epsilon_e \approx \frac{1}{207}
 ~~ \to ~ \epsilon_e  \epsilon_\mu \approx \frac{1}{3500}  
 \nonumber 
\end{align}
In a whole, these hierarchies do not exhibit  any notable regularities,  
but only feature  some order of magnitude connections. 
 Namely, by comparing the up and down quarks,  we see that 
$\epsilon_s \sim \epsilon_d$ while $\epsilon_u,\epsilon_c \sim \epsilon_d^2$.  
In other words, the up quark masses scale approximately as squares of the down quark masses. 
Comparing with leptons, we see that 
$\epsilon_\mu \sim \epsilon_d$ while $\epsilon_e \sim \epsilon_c$.   

As for the CKM matrix, the quark mixing angles are small 
(unlike the case of neutrino mixing).
Within the experimental uncertainties \cite{Workman:2022ynf}, 
they exhibit the following pattern 
\begin{align}\label{angles} 
\vert V_{us} \vert = s_{12} \approx \lambda, \quad  
\vert V_{cb} \vert = s_{23} \approx a \lambda^2, \quad 
\vert V_{ub} \vert = s_{13} \approx  2a\lambda^4 
\end{align} 
in terms of the small parameter $\lambda \approx \sqrt{1/20}$ (which is incidentally related 
as $\lambda^2 \approx \epsilon_d$ with the mass ratio  $\epsilon_d=m_d/m_s$ 
in \eqref{mass-ratios}),  and the order one numerical factor $a\approx 0.8$. 
Then for the Jarlskog invariant one has   
$J \approx 2a^2 \lambda^7\sin\delta \approx 3.5 \times 10^{-5} \sin\delta$   
which means that the smallness of observed CP-violation is originated from the 
small mixing angles rather than from a small CP-phase $\delta$, 
and in fact  $\sin\delta \simeq 1$.

\section{Fritzsch Hypothesis}
\label{Frirzsch}

It is tempting to think that 
the fermion flavour structure is connected to some underlying theory which determines
 the pattern of the Yukawa matrices with a predictive power, 
 and in particular that  the well-known formula for the Cabibbo angle 
$V_{us} = \sqrt{m_d/m_s}$ is not accidental.
Such relations between the fermion masses and mixing angles 
can be obtained by considering Yukawa matrix textures with reduced number of free parameters, 
and in particular,  by assuming that certain elements in the fermion mass matrices are vanishing 
for some symmetry reasons. 
This {\it zero-texture} approach was originally thought to calculate the Cabibbo angle in the 
two-family framework~\cite{Weinberg:1977hb,Wilczek:1977uh,Fritzsch:1977za}, 
in fact before the discovery of $b$ and $t$ quarks. 
  In the frame of six quarks, this picture  
 was extended by Harald Fritzsch \cite{Fritzsch:1977vd,Fritzsch:1979zq}  
who  suggested the following texture for the mass matrices:  
\begin{align}\label{fritzsch-3}
M_{u,d}=\left(\begin{array}{ccc}
0 & M^{u,d}_{12} & 0 \\ M^{u,d}_{21} &0 & M^{u,d}_{23} \\ 0 & M^{u,d}_{32} & M^{u,d}_{33}
\end{array}\right) 
\end{align}
where the non-zero elements are generically complex, with the symmetricity condition 
$|M^{u,d}_{ij}|=|M^{u,d}_{ji}|$   
which is motivated in the context of left-right symmetric models.\footnote{
In the original works \cite{Fritzsch:1977vd,Fritzsch:1979zq}, 
the `zeros' in these matrices were obtained  at the price of introducing several Higgs bi-doublets 
differently transforming under some discrete flavor symmetry.  
At present, this underlying theoretical construction looks rather obsolete. 
Namely, the need for several Higgs bi-doublets spoils the natural flavor conservation 
\cite{Glashow:1970gm,Glashow:1976nt,Paschos:1976ay} 
and unavoidably leads to severe flavor-changing effects \cite{Gatto:1978dy}.  
In a more natural way, 
the  Fritzsch texture for the Yukawa matrices $Y_{u,d,e}$ (rather than for the mass matrices) 
was obtained in refs. \cite{Berezhiani:1983hm,Berezhiani:1985in} in the context of models 
with horizontal $SU(3)_H$ gauge symmetry between the three fermion families 
\cite{Berezhiani:1983rk,Berezhiani:1982rr,Chkareuli:1980xnc}.  }

By rotating the phases of the upper quarks: $u^c_k \to e^{i\alpha'_k} u^c_k$ and 
$u_k \to e^{i\alpha_k} u_k$, and similarly for down quarks, the 
complex phases in the matrices \eqref{fritzsch-3} can be removed and  the non-zero entries 
 can be rendered real. 
Namely, the matrices \eqref{fritzsch-3}  can be 
parameterized as $M_{u,d} =  F'_{u,d} \widetilde{M}_{u,d} F^\ast_{u,d}$, where 
$F_u = \mathrm{diag}(e^{i\alpha_1},  e^{i\alpha_2}, e^{i\alpha_2})$ etc.
are the phase transformation matrices, and 
\begin{align}\label{fritzsch}
 F^{\prime \ast}_{u,d} M_{u,d} F_{u,d} = \widetilde{M}_{u,d}= v_w\,  \widetilde{Y}_{u,d} \, , \qquad
\widetilde{Y}_{u,d} =  \left(\begin{array}{ccc}
0 & A_{u,d} & 0 \\ A_{u,d} &0 & B_{u,d} \\ 0 & B_{u,d} & C_{u,d}
\end{array}\right) 
\end{align}
are real symmetric matrices which can  be further diagonalized 
by orthogonal transformations,  $O_{u,d}^T \widetilde{M}_{u,d} O_{u,d}= M^{\rm diag}_{u,d}$.  
The three real parameters $A_d$,  $B_d$, $C_d$ can be expressed 
in terms of the three eigenvalues of $\widetilde{Y}_d$, i.e. in terms of the down quark masses 
$m_d,m_s,m_b$.  Namely, one gets approximately, up to small corrections, 
$C_d \approx m_b/v_w$, $B_d \approx \sqrt{m_s m_b}/v_w$ and 
$A_d \approx \sqrt{m_d m_s}/v_w$, 
and similarly for up quarks and charged leptons (provided that the mass matrix of the latter 
has a structure similar to \eqref{fritzsch}).  Therefore, we have: 
\begin{align}\label{ABC} 
& C_u : B_u : A_u \approx  
1 : \epsilon_c^{1/2} : \epsilon_s \epsilon_d^{1/2}   \approx 1 : \frac{1}{16} : \frac{1}{6000}  \nonumber \\ 
& C_d : B_d : A_d \approx 
1 : \epsilon_s^{1/2} : \epsilon_s \epsilon_d^{1/2} \, \approx ~ 1 : \frac17 : \frac{1}{240}  \\
& C_e : B_e : A_e \approx 
1 : \epsilon_\mu^{1/2} : \epsilon_\mu \epsilon_e^{1/2}\,  \approx ~ 1 : \frac14 : \frac{1}{240}  \, .
\nonumber
\end{align}
Hence,  the fermion mass hierarchies \eqref{mass-ratios} 
follow from somewhat milder hierarchies between the input parameters which 
are the non-zero entries in the matrices \eqref{fritzsch}. 
Notice also the following relation between the down quark and leptons:  
\begin{align}\label{AC}
\frac{A_d}{C_d} \approx \frac{A_e}{C_e} \quad  \longrightarrow \quad 
\frac{\sqrt{m_dm_s} }{m_b} \approx \frac{\sqrt{m_em_\mu} }{m_\tau} \, .
\end{align}
Later on  we shall explore its origin in the GUT context.

The three rotation angles in the orthogonal matrix $O_d$ 
can be expressed in terms of the mass ratios $m_d/m_s$ and $m_s/m_b$. 
Analogously, the three angles in $O_u$ 
can be expressed in terms of the upper quarks mass ratios $m_u/m_c$ and $m_c/m_t$. 
The CKM matrix \eqref{CKM} is obtained as $V_\text{CKM} = O_u^T F_u^\ast F_d O_d$, 
where the diagonal matrix $F=F_u^\ast F_d$ can be parameterized by two phase parameters,  
$F=\text{diag}(e^{i\alpha}, e^{i\beta}, 1)$. Then, the four physical elements 
of the CKM matrix, that is the three mixing angles 
$\theta_{12},\theta_{23},\theta_{13}$ and the CP-phase $\delta$,  
can be expressed in terms of the four mass ratios, 
$m_d/m_s$, $m_s/m_b$, $m_u/m_c$ and $m_c/m_t$, and of two unknown phases $\beta$
and $\gamma=\beta-\alpha$. Namely, in the leading approximation one has:  
\begin{align} \label{ft-results}
\vert V_{us} \vert  
\approx \left\vert \sqrt{\frac{m_d}{m_s}} - \sqrt{\frac{m_u}{m_c}}e^{i\gamma}  \right\vert , 
\quad \vert V_{cb} \vert \approx 
\left\vert \sqrt{\frac{m_s}{m_b} }- \sqrt{\frac{m_c}{m_t} }e^{i\beta}  \right\vert , 
\quad \left\vert \frac{V_{ub}}{V_{cb}} \right\vert \approx \sqrt{\frac{m_u}{m_c}} \, .
\end{align}
Besides reproducing the formula for the Cabibbo angle, 
this texture exhibits a remarkable feature
in light of the interfamily hierarchies $m_d \ll m_s \ll m_b$ and  $m_u \ll m_c \ll m_t$.  
Namely, in the limit $A_{u,d}\to 0$ the light quarks become massless, $m_{u,d} \to 0$, 
but at the same time the first family decouples in the CKM mixing, $s_{12},s_{13} \to 0$. 
Next, in the limit $B_{u,d}\to 0$ which renders massless the second family, $m_{c,s} \to 0$, 
also its mixing with the third family disappears, $s_{23}\to 0$. 

Generalizing these properties, Fritzsch suggested that in any kind of realistic flavor models  
the quark mixing pattern should be intimately related to the interfamily hierarchy.   
The mixing angles $\theta_{ij}$ should depend on quark mass ratios 
$m_d/m_s$, $m_u/m_c$ etc. so that the smallness of the former stems from the 
smallness of the latter. 
He hypothesized the following analytic properties
\cite{Fritzsch:1983dc}: 
\begin{itemize}
\item
{\bf Decoupling hypothesis: }  
in the limit  of massless first family, $m_u,m_d \to 0$,  
its mixings with the heavier families disappear, i.e. $\theta_{12}, \theta_{13} \to 0$. 
At the next step, for massless second family, $m_s,m_c \to 0$, 
also the 2-3 mixing should disappear, i.e. $\theta_{23} \to 0$. 
\vspace{2mm}
\item
{\bf Scaling hypothesis: }  in the limit when 
up and down quark masses become proportional, $m_u : m_c : m_t = m_d : m_s : m_b$, 
all mixing angles should vanish: $\theta_{12},\theta_{23},\theta_{13} \to 0$.  
\end{itemize} 

While the Fritzsch ansatz \eqref{fritzsch} has the first property ({\it decoupling} hypothesis), 
in general it does not meet the second one (scaling hypothesis), since 
the Yukawa matrices $Y_u$ and $Y_d$ do not necessarily become aligned 
in the limit $A_u:B_u:C_u = A_d:B_d:C_d$. In fact, in this case we have 
$m_u : m_c : m_t = m_d : m_s : m_b$ but the mixing angles 
do not generally vanish because of the arbitrary phases $\beta$ and $\gamma$.

However, the original Fritzsch texture for quarks was excluded when 
the knowledge of quark masses and CKM parameters became accurate enough
\cite{Kang:1992eq}.  
Given the present experimental and lattice results on quark masses and CKM elements, 
there is no parameter space  in which the precision data can be reproduced.
More concretely, the small enough value of $|V_{cb}|$ and large enough value of $|V_{ub}/V_{cb}|$ 
cannot be achieved for any values of the phase parameters $\beta$ and $\gamma$ in \eqref{ft-results}. 
A possibility to obtain viable textures  
is to extend the original Fritzsch texture by replacing one of the zero entries 
with a non-zero one, e.g. by introducing a non-zero 13 element \cite{Berezhiani:1991tj} 
or a non-zero 22 element, as e.g. in refs. \cite{Xing:2015sva,Fritzsch:2021ipb} 
(for a review of different schemes, see \cite{Fritzsch:1999ee}). 
However,  the introduction of new parameters reduces the predictive power, 
and, in addition, these modifications generically do not satisfy  the {\it decoupling} feature.   

On the other hand, instead of decreasing the number of zero entries,
one can think to break the symmetricity condition.
Namely, an asymmetry in the 23 blocks, $|M^{u,d}_{23}| \neq |M^{u,d}_{32}|$ 
can be introduced in the mass matrices 
\cite{Berezhiani:1996ii,Berezhiani:1998vn,Berezhiani:2000cg,Roberts:2001zy}.
In other words, one can consider  the Yukawa textures of the form 
\begin{align}\label{fritzsch-mod}
Y_f = F'_f \widetilde{Y}_{f} F_f \, , \qquad \widetilde{Y}_{f} =  \left(\begin{array}{ccc}
0 & A_{f} & 0 \\ A_{f} &0 & x_f B_{f} \\ 0 & x_f^{-1} B_{f} & C_{f} \end{array}\right), 
\quad   f=u,d,e
\end{align}
where the matrices $F,F'$ contain only phases and 
$\widetilde{Y}$ are the real matrices, with $x_{u,d,e}$ being deformation parameters. 
It is worth noting that such a modification preserves the main properties of the original Fritzsch 
texture. In particular, it satisfies the {\it decoupling} property, and in addition  
the hierarchy relations \eqref{ABC} still remain valid. 

As it was shown in a recent analysis in ref. \cite{Belfatto:2023qca}, such a texture 
is perfectly compatible with the observed pattern of the CKM matrix. 
Namely, all defects of the original texture can be corrected at once provided that 
the upper quark matrix exactly has a `symmetric'  Fritzsch texture, that is $x_u=1$, 
and only the down quark matrix is deformed by a factor $x_d = 3$  or so. 
In fact, such a modification leads to the correct values of the CKM mixing angles as well as of the CP-violating phase as functions of the fermion mass ratios and of the two phase factors  \cite{Belfatto:2023qca}.

In next section 
we describe how the Fritzsch texture can be obtained within the context of 
the inter-family gauge group $SU(3)_H $, and how it can be minimally deformed in the 2-3 blocks  
by using a scalar field in adjoint (octet) representation of $SU(3)_H$. 
In section \ref{texture}  
we analyse the predictions of Fritzsch textures  in light of the present high precision 
determinations of quark masses and CKM matrix elements
and show that this flavour structure predicts 
 the mixing angles and the CP-violating phase in perfect agreement 
with the experimental results.
 
 \section{Fritzsch-like textures from horizontal symmetry $\mathbf{SU(3)_H} $ }
\label{model}

The key for understanding the replication of families, the fermion mass hierarchy and the mixing pattern 
may lie in symmetry principles. 
For example, one can assign to the fermion species different charges of an abelian global flavor symmetry  
$U(1)$  \cite{Froggatt:1978nt}.  
There are also models making use of an anomalous gauge symmetry $U(1)_A$ 
to explain the fermion mass hierarchy while also tackling other naturalness issues 
\cite{Ibanez:1994ig,Binetruy:1996xk,Dudas:1996fe,Berezhiani:1996nu,Berezhiani:1996bv}. 
However, it is difficult to obtain the highly predictive
quark mass matrices with the texture zeros within this approach. 

One can point to a more complete picture by introducing a non-abelian horizontal 
gauge symmetry $SU(3)_H$ between the three families  
\cite{Berezhiani:1983hm,Berezhiani:1985in,Berezhiani:1983rk,Berezhiani:1982rr,Berezhiani:1990wn}. 
This symmetry should have a chiral character, 
with the LH and RH components of quarks (and leptons) transforming 
in different representations of the family symmetry. 
In particular, they can be arranged respectively as  the 
triplet and anti-triplet representations of $SU(3)_H$ which  
in our notation means that LH fermions $q,\ell$ as well as anti-fermions 
$u^c,d^c,e^c$ are  $SU(3)_H$ triplets:
\begin{align}\label{SM-triplets}
q_i, \, u^c_i , d^c_i \sim 3 \, ; \quad \quad \ell_i, \, e^c_i \sim 3 
\end{align} 
with $i=1,2,3$ being the family $SU(3)_H$ index. Clearly, in this case the fermion direct 
Yukawa couplings to the Higgs doublet is forbidden by $SU(3)_H$,  
which means that their masses cannot be induced 
without the breakdown of the horizontal symmetry. 
On the contrary, a vector-like $SU(3)_H$, 
with $q,\ell \sim 3$ and $u^c,d^c,e^c \sim \bar3$, would allow 
the Yukawa couplings  inducing  a degenerate fermion 
mass spectrum between the three families, without breaking $SU(3)_H$. 

In addition, the vector-like $SU(3)_H$ would not allow the grand unified 
extensions of the SM as $SU(5)$ \cite{Georgi:1974sy} or $SO(10)$ \cite{Fritzsch:1974nn}, 
while the chiral arrangement \eqref{SM-triplets} 
is compatible with the GUT structures (see e.g. \cite{Berezhiani:2005ek} 
for a review on fermion patterns in GUTs).  
In particular, in $SU(5)$  each family is represented by the LH spinors   
in $\bar5$ and $10$ representations. 
Thus, in the context of $SU(5)\times SU(3)_H$ all fermion species in \eqref{SM-triplets} 
are embedded in the following representations 
\cite{Berezhiani:1985in,Berezhiani:1983rk,Berezhiani:1982rr}:\footnote{$SU(5) \times SU(3)_H$ 
can be embedded into $SU(8)$ group \cite{Chkareuli:1980xnc}  
but this possibility introduces in the particle spectrum  some extra  fermions in exotic representations.}
\begin{align}\label{SU5-triplets}
\bar{F}_i = (d^c, \, \ell)_i \sim (\bar5,3) , \quad \quad T_i = (u^c, \, q, \, e^c)_i \sim (10,3)  \, .
\end{align} 
As for $SO(10)$ GUT, all fermions of one family can be packed 
into the 16-dimensional spinor representation of $SO(10)$, $\Psi = (\bar{F}, T, \nu^c)$, 
along with the ``right-handed neutrino" $\nu^c_{L}=C\overline{\nu_R}^T$.   
Hence, in the context of $SO(10)\times SU(3)_H$ all three families compose  
the unique multiplet  $\Psi_i = (\bar{F}, T, \nu^c)_i \sim (16,3)$.\footnote{More generically,
in the SM context  the maximal chiral symmetry which can be achieved in the limit 
of vanishing Yukawa couplings \eqref{Yukawas-SM} is $U(3)^5$ 
independently  transforming the different species $q,u^c,d^c,\ell,e^c$. 
In the context of $SU(5)$, the maximal flavor symmetry is reduced to $U(3)^2= 
U(3)_{\bar F} \times U(3)_T$ independently transforming fermionic 
$\bar 5$- and 10-plets, while the $SO(10)$ structure 
allows a unique flavor symmetry $U(3)_H$ from which we gauge a non-abelian 
$SU(3)_H$ part, with chiral species arranged as in \eqref{SM-triplets}.  
} 
With this set of fermions,  $SU(3)_H$ would have triangle anomalies. 
For their cancellation one can introduce additional  chiral fermions which are 
SM singlets but transform nontrivially under $SU(3)_H$ 
 \cite{Berezhiani:1983hm,Berezhiani:1985in,Berezhiani:1983rk}, and they can be used 
 for neutrino mass generation \cite{Berezhiani:1983obe,Berezhiani:1985vxy,Anselm:1985we}. 
The easiest way to cancel the anomalies, suggested in ref. \cite{Berezhiani:1996ii}, 
is to share the $SU(3)_H$ symmetry with a parallel mirror sector of particles 
having exactly the same physics as ordinary particles  (for a review, see e.g. 
\cite{Berezhiani:2003xm,Berezhiani:2005ek,Berezhiani:2008zza}). 
In this case $SU(3)_H$ anomalies will be cancelled between ordinary fermion species 
and their mirror counterparts of opposite chiralities.

The chiral character of the fermion representations  \eqref{SM-triplets} 
forbids their direct Yukawa couplings of with the Higgs doublet $\phi$, 
so that the fermion masses cannot be induced without  breaking $SU(3)_H$.  
As far as the fermion bilinears $u^c_i q_j$, $d^c_i q_j$ and $e^c \ell$ 
transform in representations $3\times 3= 6 + \bar3$,  
the fermion masses can be induced only via the higher order operators 
\begin{align}\label{operators}
\sum_n \left(C^{(u)}_n \frac{\, \chi^{ij}_n }{M} \, u^c_i q_j  \phi \, + \, 
C^{(d)}_n  \frac{\, \chi^{ij}_n }{M} \, d^c_i q_j \tilde\phi  
\, + \, C^{(e)}_n  \frac{\, \chi^{ij} _n}{M} \, e^c_i \ell_j \tilde\phi \right) ~ + ~ \text{h.c.}
\end{align}
involving some amount $n$ of horizontal scalars $\chi_n$ (coined as flavons) in symmetric 
(anti-sextets $\chi^{\{ij\}} \sim \bar6$) or antisymmetric 
(triplets $\chi^{[ij]} =\epsilon^{ijk}\chi_k \sim 3$) representations of $SU(3)_H$ 
which are gauge singlets of the SM. 
Here $M$ is some effective scale and the Wilson coefficients $C^{(f)}_n$ ($f=u,d,e$) 
are generically complex.    
After inserting the flavon VEVs, the operators in eq. \eqref{operators} give rise to 
the SM Yukawa couplings  \eqref{Yukawas-SM} as  
$Y^{ij}_f=\sum_n C^{(f)}_n \langle \chi^{ij}_n \rangle /M$.  
In a sense, these operators ``project"  the VEV pattern of the flavon fields $\chi$ 
onto the structure of the Yukawa matrices.  
In the UV-complete picture such operators can be induced in a seesaw-like manner, 
via integrating out some extra heavy scalars  \cite{Berezhiani:1983rk,Berezhiani:1982rr} 
or extra heavy fermions in vector-like representations 
\cite{Berezhiani:1983hm,Berezhiani:1985in} 
living at the mass scale $\sim M$.

Interestingly,  besides being invariant under the local 
$SU(3)_H$ symmetry by construction, these operators \eqref{operators} exhibit in fact a larger global symmetry 
$U(3)_H=SU(3)_H \times U(1)_H$. 
Namely,  they are invariant also under an accidental  global chiral $U(1)_H$ symmetry,  
implying the overall phase transformation of fermions 
$(u^c_i, d^c_i,q_i )\to e^{i\omega} (u^c_i, d^c_i,q_i )$ 
and flavon scalars $\chi_n \to e^{-2i\omega} \chi_n$.
Hence, all families can become massive only if $U(3)_H$ symmetry is fully broken. 

 This feature allows to relate the fermion mass hierarchy and mixing pattern with the 
breaking steps of $U(3)_H$ symmetry, with a natural realization of the {\it decoupling} hypothesis.  
When $U(3)_H$ breaks down to $U(2)_H$, the fermions of the third fermion family get masses 
while the first two families remain massless and  all mixing angles are vanishing.
At the next step, when $U(2)_H$ breaks down to $U(1)_H$,  
the second family acquires masses and the CKM mixing angle $\theta_{23}$ can be non-zero, 
but the first family remains massless ($m_u,m_d=0$) and unmixed with the heavier 
fermions  ($\theta_{12},\theta_{13}=0$). Only at the last step, when $U(1)_H$ is broken, 
also the first family can acquire masses and its mixing with heavier families can emerge. 
In this way, the inter-family mass hierarchy can be related to the hierarchy  
of flavon VEVs inducing the horizontal symmetry breaking 
$U(3)_H\to U(2)_H \to U(1)_H \to$ {\it nothing}.

In the last step of this breaking chain, the chiral global $U(1)_H$ symmetry 
can be associated with the Peccei-Quinn symmetry provided that $U(1)_H$ is also respected 
by the Lagrangian of the flavon fields \cite{Berezhiani:1985in,Berezhiani:1990wn}. 
This can be achieved by forbidding 
the trilinear terms between the $\chi$-scalars by means of a discrete symmetry. 
Thus, in this framework, the Peccei-Quinn symmetry can be considered as 
an accidental symmetry emerging from the local symmetry $SU(3)_H$. 
In this case the axion will have non-diagonal couplings between  
the fermions of different families \cite{Berezhiani:1983hm,Berezhiani:1985in,Berezhiani:1990wn}.  
Phenomenological and cosmological implications of gauge family symmetry 
with such flavor-changing axion were discussed in refs. 
\cite{Berezhiani:1989fp,Berezhiani:1990jj,Berezhiani:1990sy,Berezhiani:1989fs,Berezhiani:1989fu,Berezhiani:1992rk}. 

In scenarios with horizontal $SU(3)_H$ symmetry  the Fritzsch textures can be naturally obtained 
by a suitable choice of  the representations and VEV configurations of the $\chi$-flavons.  
As the simplest set,  one can take two triplets $\chi_1$, $\chi_2$,
and one anti-sextet $\chi_3$, with their VEVs in the following form \cite{Berezhiani:1985in}: 
\begin{align}\label{vevs}
&\langle\chi^{\{ij\}}_3 \rangle=\text{diag}(0,0,V_3) \qquad
\langle\chi_{2i} \rangle=\left(\begin{array}{c} V_2 \\ 0 \\0 \end{array}\right) \qquad
\langle\chi_{1i}\rangle=\left(\begin{array}{c} 0 \\ 0 \\V_1 \end{array}\right) \, ,
\end{align}
i.e. the VEV of $\chi_3 $ is given by a symmetric rank-1  matrix 
which for a convenience can be directed  towards the 3rd axis in the $SU(3)_H$ space. 
As for the VEVs of $\chi_1$ and $\chi_2$, they are respectively parallel and orthogonal 
to $\langle \chi_3 \rangle$ (so that $\langle \chi_3 \rangle$ can be oriented towards the 3rd axis 
 without losing generality; the detailed analysis of the flavon scalar potential 
is given in ref.~\cite{Berezhiani:1983rk}). 
Thus, the total matrix of flavon VEVs has the Fritzsch  texture: 
\begin{align}\label{VEV-matrix}
\langle \chi^{ij} \rangle = \left\langle \chi_1^{[ij]} + \chi_2^{[ij]} + \chi_3^{\{ij\}} \right\rangle= 
  \left(\begin{array}{ccc} 0 &  V_1 & 0 \\ -  V_1 & 0 &   V_2 \\ 0 & -  V_2 &  V_3 
  \end{array}\right) \, .
\end{align}
After the $SU(3)_H$ breaking, the theory reduces to the SM containing 
one standard Higgs doublet $\phi$,  and the `projective' operators \eqref{operators} 
transfer the Fritzsch-like pattern of the VEV matrix \eqref{VEV-matrix}
 to the Yukawa matrices $Y_f = \sum_n C^{(f)}_n \langle \chi_n \rangle/M$ ~($f=u,d,e$), 
modulo  different constants 
$C^{(u)}_n$, $C^{(d)}_n $ and $C^{(e)}_n$ in the expansion \eqref{operators}.\footnote{
Thus, in this construction the flavor is naturally conserved in neutral currents, 
in difference from the Fritzsch's original model \cite{Fritzsch:1977vd,Fritzsch:1979zq}
involved several Higgs doublets.  } 
(The $-$ signs between off-diagonal terms are irrelevant since they 
can be eliminated by quark phase transformations.)  
Namely,  after rotating away the complex phases, the Yukawa matrices acquire
the `symmetric' Fritzsch forms $\widetilde{Y_f}$ as in eq. \eqref{fritzsch}.  
 The inter-family mass hierarchies can be related to a hierarchy 
in the horizontal symmetry breaking chain 
$U(3)_H\to U(2)_H \to U(1)_H \to$ {\it nothing}, that is $V_3 \gg V_2 \gg V_1$. 
In particular, the hierarchies \eqref{ABC} between the Yukawa entries can 
be originated from the hierarchy of the flavon VEVs which can be estimated as 
\begin{align}\label{VEV123}
V_3 : V_2 : V_1 \simeq 
1 : 1/5 : 1/100 ~ \Longrightarrow ~ 
U(3)_H \stackrel{V_3}{\to} U(2)_H \stackrel{V_2}{\to} U(1)_H \stackrel{V_1}{\to}  \mathrm{nothing}  \, .
\end{align} 
 
 In the context of $SU(5)\times SU(3)_H$ theory  
\cite{Berezhiani:1982rr,Berezhiani:1983rk,Berezhiani:1985in}, 
with the quarks and leptons in representations \eqref{SU5-triplets},  
one can consider the $SU(5)$-invariant effective operators
\begin{align}
\label{operators-SU5}
\sum_n \left( C^{(10)}_n \frac{\, \chi^{ij}_n }{M} \,  T_i T_j  H   + 
C^{(5)}_n \frac{\, \chi^{ij}_n }{M} \,  \bar{F}_i T_j \bar{H} \right) 
 \, + \, {\rm h.c.}  
\end{align} 
where $H$ is the scalar $5$-plet which contains the SM Higgs doublet $\phi$. 
In fact, after $SU(5)$ breaking down to the SM, the last two terms in \eqref{operators}
emerge from the decomposition of the last  term in \eqref{operators-SU5}. 
Then, if all flavons $\chi_n$ are $SU(5)$ singlets, 
the Yukawa constants of down quarks and charged leptons would be identical, $Y_e = Y_d^T$.  
Though $b-\tau$ unification is a successful prediction of $SU(5)$, for the first two families 
this  would imply the wrong relations $m_b : m_s : m_d = m_\tau : m_\mu : m_e$, 
in obvious contradiction with the observed pattern  given in \eqref{mass-ratios}. 
In addition, in the first term of \eqref{operators-SU5}  the antisymmetric flavons 
cannot contribute by symmetry. Hence, one has $A_u,B_u = 0$ and so $u$ and $c$ quarks 
remain massless.
However, this shortcoming can be avoided by assuming that some (at least one)  
of the flavon triplets transforms as 24-plet of $SU(5)$  
\cite{Berezhiani:1985in}.\footnote{More generally, the effective coefficients $C_n$ 
in \eqref{operators-SU5} 
can be functions of  a scalar 24-plet $\Sigma$ which breaks $SU(5)$ down to 
the SM $SU(3)\times SU(2)\times U(1)$ \cite{Berezhiani:1995tr}.}    
This can avoid the wrong relation $Y_e = Y_d^T$  
between the down quark and lepton Yukawas, 
and also induce the non-zero $A_u$ and $B_u$ in $Y_u$.

\begin{figure}
\centering
\includegraphics[width=0.48\linewidth]{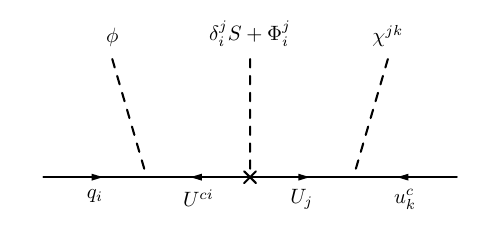}
\hspace{10pt}
\includegraphics[width=0.48\linewidth]{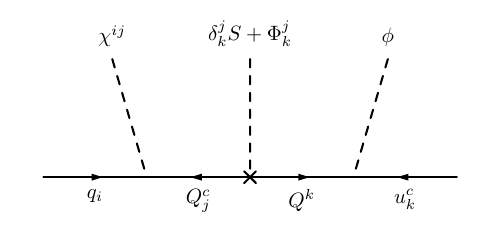}
\caption{Seesaw diagrams inducing the Yukawa couplings of upper quarks 
via exchange of vector-like quarks $U,U^c$ and $Q,Q^c$. 
Analogous diagrams with $u^c\to d^c$, $U,U^c\to D,D^c$ and $\phi \to \tilde\phi$ will work for down quarks.} \label{massdiag} 
\end{figure}

Let us turn back to the general case of the SM, 
with the quark and lepton species in $SU(3)_H$ representations \eqref{SM-triplets}.
In the UV-complete pictures the operators \eqref{operators}  
can be induced via renormalizable interactions after integrating out  some extra heavy fields, 
scalars \cite{Berezhiani:1983rk,Berezhiani:1982rr},  or verctor-like fermions 
\cite{Berezhiani:1983hm,Berezhiani:1985in} (see also \cite{Berezhiani:1990wn}). 
The latter possibility is more economic and natural, 
and we shall exploit it in our further considerations.\footnote{
In particular, in the context of supersymmetric theories with $SU(3)_H$ family symmetry  
this mechanism can lead to interesting relations 
between the fermion Yukawa couplings and the soft SUSY breaking terms which  
naturally  realize the minimal flavor violation scenarios 
\cite{Berezhiani:1996ii,Berezhiani:1996kk,Anselm:1996jm,Berezhiani:2001mh}.  
} 
Namely,  one can introduce the following vector-like set of chiral LH fermions 
of the up- and down-quark type in weak doublet and singlet representations
 \begin{align}\label{VL}
Q^i = ({\mathcal U},{\mathcal D})^i , \,  U^{ci},\, D^{ci} \sim \bar3  \qquad  
Q^c_i = ({\mathcal U}^c,{\mathcal D}^c)_i , \, U_i \,  D_i \sim 3   \, .
\end{align} 
 Then the matrices of Yukawa couplings $Y_{u,d}$ in eq.  \eqref{Yukawas-SM} are induced 
after integrating out these fermion species with large mass terms.
 Analogously, the charged lepton Yukawa couplings 
can be induced by using the vector like lepton states, weak singlets 
 $E_i \sim 3$, $E^{ci} \sim \bar3$ and weak doublets $L^i \sim \bar3$, $L^c_i \sim 3$. 
 In particular, for the upper quarks this seesaw-like mechanism  is illustrated by two diagrams 
in fig.~\ref{massdiag} while the analogous diagrams 
will work for the down quarks and leptons.
For the heavy vector-like species  having $SU(3)_H$ invariant masses, this mechanism 
will induce the operators \eqref{operators} 
which `project'  the Fritzsch-texture \eqref{VEV-matrix}  of the flavon VEVs 
on the quark and lepton  Yukawa matrices. 
 
However, by using the diagrams of fig.~\ref{massdiag},   
one can also achieve a realistic deformation of the Fritzsch ansatz in the following way.  
Let us assume that the the masses of vector-like fermions \eqref{VL} are 
induced by spontaneous symmetry breaking from the VEVs of some scalars.  
Their mass terms transform as $\bar3\times 3=1+8$ and so one can introduce 
the Yukawa couplings of $D,D^c$ fermions  
\begin{align}\label{Yuk-Phi} 
(g_D S \delta_j^i +  f_D \Phi_j^i) D_i D^{cj}  + {\rm h.c.} 
\end{align}
and analogous couplings of $U, U^{c}$ and $Q^c,Q$ species,
with the scalars in singlet and octet representations of $SU(3)_H$, 
$S \sim 1$ and $\Phi \sim 8$. 
In fact,  one can introduce an adjoint scalar $\Phi$ of $SU(3)_H$ 
in analogy to the adjoint scalar of $SU(5)$, the 24-plet $\Sigma$,  
which possibility was discussed in refs. \cite{Berezhiani:1996ii,Berezhiani:2000cg}. 
 We also assume that the cross-interaction terms of $\Phi$ with $\chi$-flavons in the 
 scalar potential align the VEV of $\Phi$  towards the largest VEV $V_3$ 
 in \eqref{VEV-matrix},
i.e. the VEV is proportional to the $\lambda_8$ generator of $SU(3)_H$: 
$\langle \Phi \rangle = V \, \mathrm{diag}(1,1,-2)$, and it breaks the $SU(3)_H$  symmetry 
down to $SU(2)_H \times U(1)_H$. 
Then, the masses of the heavy species are contributed by both the singlet and the octet VEVs 
and their mass matrices have diagonal forms, namely  
$\mathrm{M}_{D} = g_{D} \langle S\rangle + f_{D} \langle \Phi \rangle = 
M_{D} \, \mathrm{diag} (X_{D},X_D,1)$, and analogously for $U$ and $Q$ species,  
where generically $X_{U,D,Q} \neq 1$ are complex numbers. 
In this way, one obtains the Yukawa matrices with the 
minimally deformed Fitzsch textures \eqref{fritzsch-mod},  
where the `asymmetries'  $x_{u,d}$ are  introduced between 23 and 32 elements of 
$\widetilde{Y}_{u,d}$, without changing the amount of zero elements.  
 
In the case of $SU(5)\times SU(3)_H$ \cite{Berezhiani:1985in}, 
with ordinary quarks and leptons  arranged as in \eqref{SU5-triplets}, 
the vector-like fermions can be introduced in the representations
\begin{align}\label{SU53} 
& \mathcal{T}^i = (U^{ci},Q^i,E^{ci}) \sim (10,\bar3) , 
\qquad \overline{\mathcal{T}}_i  = (U_i, Q^c_i, E_i)  \sim  (\overline{10},3) \nonumber \\ 
& \quad \overline{\mathcal{F}}^i =  (D^{ci}, L^i) \sim (\bar5,\bar3) , \qquad \qquad
\mathcal{F}_i = (D_i,L^c_i) \sim (5,3) 
\end{align}
which include all vector-like quarks in \eqref{VL}  together with lepton states $L,L^c$ and $E^c,E$.  
  
In principle, all Yukawa couplings  $Y_{u,d,e}$ can be induced by diagrams 
similar to that of fig.~\ref{massdiag} involving $\mathcal{T},\overline{\mathcal{T}}$ states,  
while $Y_{d,e}$ can be contributed also by exchange of 
$\mathcal{F},\overline{\mathcal{F}}$ states \cite{Berezhiani:1985in}. 
Using these tools, in ref. \cite{Belfatto:2023qca} the model was built which 
gives the Yukawa matrices \eqref{fritzsch-mod} in a particularly constrained form.  
Namely, the (de-phased) up quark Yukawas is predicted 
to have a `symmetric' Fritzsch texture, $x_u=1$:  
\begin{align}\label{Yu} 
\widetilde{Y}_{u} =  \left(\begin{array}{ccc}
0 & A_{u} & 0 \\ A_{u} &0 &  B_{u} \\ 0 & B_{u} & C_{u} \end{array}\right) 
\end{align}
while for the down quarks and leptons one has deformed patterns with $x_{d,e}\neq 1$: 
\begin{align}\label{Yde} 
\widetilde{Y}_{d} =  \left(\begin{array}{ccc}
0 & A_{d} & 0 \\ A_{d} &0 &  x_d B_{d} \\ 0 & x_d^{-1} B_{d} & C_{d} \end{array}\right), 
\qquad 
\widetilde{Y}_{e}^T =  \left(\begin{array}{ccc}
0 & A_{e} & 0 \\ A_{e} &0 &  x_e B_{e} \\ 0 & x_e^{-1} B_{e} & C_{e} \end{array}\right) 
\end{align}
We shall not describe the model of ref. \cite{Belfatto:2023qca} in details here, 
but only briefly mention its key ingredientts: \\
(i) flavon fields $\chi_1$ and $\chi_3$ are taken as $SU(5)$ singlets, 
i.e. $\chi_1 \sim (1, 3)$ and $\chi_3 \sim (1,\bar6)$, while $\chi_2$ is taken 
in a mixed representation $\chi_2 \sim (24,3)$, with the horizontal directions  of the VEVs 
 as in \eqref{vevs} with hierarchy as in \eqref{VEV123} between their values. 
This `mixed'  choice of $\chi_2$ induces non-zero $B_u$ in \eqref{Yu}. It also 
introduces $SU(5)$ breaking Clebsch factors in 23 blocks of \eqref{Yde}, 
so that $B_d\neq B_e$,     
thus avoiding the wrong scaling relation $m_b : m_s : m_d = m_\tau : m_\mu : m_e$ 
between down quark and lepton masses.  But $SU(5)$ invariance   
for the other two entries is kept, i.e. $C_d=C_e$ and $A_d=A_e$, which in turn 
implies the successful relation \eqref{AC}. \\
(ii) heavy 10-plets $\mathcal{T},\overline{\mathcal{T}}$ get the dominant mass contribution from 
the Yukawa coupling with a field $S$ which is singlet of $SU(5)\times SU(3)_H$, 
with a subdominant contribution from the $SU(5)$ adjoint Higgs $\Sigma$. This explains 
the smallness of $A_u/C_u$ with respect to $A_d/C_d$ (see eq. \eqref{ABC}; 
in fact, without the latter contribution of $\Sigma$, we would have $A_u =0$; \\
(iii) the masses of heavy 5-plets $\mathcal{F},\overline{\mathcal{F}}$ are strongly contributed 
by the Yukawa couplings with the $SU(3)_H$ adjoint scalar $\Phi$ which induces 
the deformation factors $x_{d,e} \neq 1$. 
  
In the next section we mainly concentrate on the implications of the textures in \eqref{Yu}
and \eqref{Yde}  for the quark sector, following  the analysis performed 
in ref. \cite{Belfatto:2023qca}, 
but having in mind that in the context of grand unification
analogous considerations can be extended to leptons.

\section{Deformed  Fritzsch textures: CKM matrix vs. quark mass ratios}
\label{texture}

Let us consider the Yukawa matrices $\mathrm{Y}_{u,d}$ having the form \eqref{fritzsch-mod}. 
They can be  brought to the diagonal forms 
 \begin{align}\label{yukawa-eigenvalues}
 Y^\text{diag}_u = \text{diag}(y_u,y_c,y_t) , \qquad 
Y^\text{diag}_d = \text{diag}(y_d,y_s,y_b) 
\end{align}
via  biunitary transformations as in eq. \eqref{diag}  with the unitary matrices 
$V_u = F_u O_u$ and $V_d = F_d O_d$, 
where $F_{u,d}$ are the diagonal matrices containing phases, and 
$O_{u,d}$ are the real orthogonal matrices. 
In other terms, by phase transformations  $F'_{u,d} Y_{u,d} F_{u,d}$ with  
$F_{u,d}=\text{diag}(e^{i\alpha_{u,d}}, e^{i\beta_{u,d}}, e^{i\gamma_{u,d}})$ (and similarly $F'_{u,d}$),  
the Yukawa matrices $Y_{u,d}$ are brought to the real forms $\widetilde{Y}_{u,d}$   
which can be further diagonalized by bi-orthogonal transformations 
$O'_{u,d}  \widetilde{Y}_{u,d} O_{u,d} = Y^\text{diag}_{u,d}$. 

 Thus, for the CKM matrix of quark mixing we obtain
\begin{align}\label{CKMF}
V_\mathrm{CKM} = V_u^\dag V_d=O_u^T F O_d =  \:
O_u^T
\left(\begin{array}{ccc}
e^{i(\tilde{\beta}+\tilde{\delta})}& 0& 0 \\ 0 & e^{i \tilde{\beta}} &0 \\0&0&1
\end{array}\right)
O_d  \, ,
\end{align}
where the matrix $F=F_u^\dag F_d$ without loss of generality 
can be parameterized by the two phases $\tilde\beta$ and $\tilde\delta$ 
while the orthogonal matrices $O_{u,d}$ can be parametrized  as
\begin{align}\label{Od}
& O_d=O_{d23}O_{d13}O_{d12}= \left(\begin{array}{ccc}
 1 & 0    & 0 \\
 0 & c^d_{23} & s^d_{23} \\
 0 & -s^d_{23} & c^d_{23}
\end{array}\right) 
\left(\begin{array}{ccc}
c^d_{13} & 0 & s^d_{13} \\
0    & 1 & 0 \\
-s^{d}_{13}  & 0 & c^d_{13}
\end{array}\right)
\left(\begin{array}{ccc}
c^d_{12} &   s^d_{12} & 0 \\
 -s^d_{12}   & c^d_{12} & 0  \\
 0 & 0 &1
\end{array}\right) \, ,
\end{align}
with $c^d_{ij}=\cos\theta^d_{ij}$ and $s^d_{ij}=\sin\theta^d_{ij}$.
Analogously, for up-quarks we have $O_u=O_{u23}O_{u13}O_{u12}$, as well as for
the rotations of right-handed states $O'_{u,d}$.  

Hence, $\widetilde{Y}_d$ contains four parameters, $A_d,B_d,C_d$ and $x_d$, 
which determine the three Yukawa eigenvalues $y_{d,s,b}$ and the three rotation angles in $O_d$. 
Analogously, the four parameters in $\widetilde{Y}_u$ determine the Yukawa eigenvalues 
$y_{u,c,t}$ and the three angles in $O_u$. 
Therefore, we have $10$ real parameters $A_{u,d},B_{u,d},C_{u,d},x_{u,d}$ 
and two phases $\tilde{\beta},\tilde{\delta}$ which 
have to match $10$ observables, the $6$ Yukawa eigenvalues   
and $4$ independent parameters of the CKM matrix (see eq. \eqref{ckmSP}).

The Yukawa eigenvalues and rotation matrices $O$ and $O'$ can be found by 
considering the ``symmetric" squares respectively of the Yukawa matrices $\widetilde{Y}_f^T \widetilde{Y}_f$ and 
$\widetilde{Y}_f\widetilde{Y}_f^T$, $f=u,d$. 
In doing so, we obtain the following relations 
\begin{align}\label{secular} 
&  C^2 +  (x^2+x^{-2})B^2 + 2A^2   = Y_3^2 + Y_2^2 + Y_1^2 
  \nonumber \\
&  B^4 + 2C^2 A^2  + (x^2+x^{-2})B^2A^2 + A^4 
 = Y_3^2 Y_2^2 +  Y_3^2 Y_1^2 + Y_2^2 Y_1^2 \nonumber \\
&  
A^2 C = Y_1 Y_2 Y_3 
\end{align}
where we omit the indices $f=u,d$ and imply $Y_{1,2,3}=y_{u,c,t}$ for the Yukawa 
eigenvalues of upper quarks and $Y_{1,2,3}=y_{d,s,b}$ for down quarks. 

 Having in mind the  relations  \eqref{mass-ratios}, we can use them 
 in the synthetic form $y_t : y_c : y_u\sim 1 : \epsilon_u : \epsilon_u^2$ and
$y_b : y_s : y_d\sim 1 : \epsilon_d : \epsilon_d^2$,  
noting that phenomenologically $\epsilon_u \sim \epsilon_d^2$.
Then, in leading approximation (up to relative corrections of order 
$\epsilon \sim Y_2/Y_3\sim Y_1/Y_2$) we have \cite{Berezhiani:1998vn,Belfatto:2023qca}: 
\begin{align}\label{leading}
C\approx Y_3 , \quad B \approx \sqrt{Y_2Y_3}, \quad A \approx \sqrt{Y_1Y_2}
\end{align}
so that $C_f : B_f : A_f \sim 1 : \epsilon_f^{1/2} : \epsilon_f^{3/2}$. 
Since the ratios $C_d: B_d : A_d \sim V_3 : V_2 :V_1$  
in fact reflect the hierarchy in the horizontal symmetry breaking \eqref{VEV123}
(and also taking into account  that the off-diagonal elements in $Y_u$  
have additional suppression in the context of the $SU(5)\times SU(3)_H$ model \cite{Belfatto:2023qca}),  
this means that the inter-family mass hierarchy 
 can actually be induced by a milder hierarchy between the flavon VEVs. 
The matrix entries $A_f$, $B_f$ and $C_f$ depend on the deformation $x_{f}$ 
only at higher orders in $\epsilon_{f}$.
 As regards the rotation matrices, the angles in \eqref{Od},  
up to small corrections of order $\epsilon_d$,  are given by
\begin{align}\label{sd}
 s^{d}_{23} \approx \frac{1}{x_d}\sqrt{\frac{y_s}{y_b} }  , \quad 
 s^{d}_{12}  \approx \sqrt{\frac{y_d}{y_s} } , \quad 
 s^{d}_{13} \approx 
x_d\, \frac{y_s}{y_b}\sqrt{\frac{y_d}{y_b}} \, .
\end{align} 
 The expressions for $s^\prime_{d23}$, $s^{\prime}_{d12}$ and $s^{\prime}_{d13}$ 
are the same with the replacement $x_d\rightarrow x_d^{-1}$,
Analogously, the up quark rotation angles in $O_u$ can be expressed in terms 
of Yukawa ratios $y_u/y_c$ and $y_c/y_t$.  
Then,  at leading order, for the elements of we obtain: 
\begin{align}\label{vamanoeps}
& |V_{us}| \approx \left|\, s_{12}^d-   s_{12}^u \, e^{i\tilde{\delta}}\, \right| \approx |V_{cd}|\, , 
 \qquad 
|V_{cb}| \approx \left|\, s_{23}^d - s_{23}^u \,   e^{i \tilde{\beta}}\, \right| \approx |V_{ts}| \\ 
 & |V_{ub}| \approx \left| s_{13}^d\, e^{i\tilde{\delta}} -s_{12}^u \,  
\left(  s_{23}^d - s_{23}^u  \,e^{-i\tilde{\beta}}\, \right)  \right|\, , \quad 
|V_{td}| \approx \left| s_{13}^d\, e^{i\tilde{\delta}} -s_{12}^d \, 
 \left( s_{23}^d  - s_{23}^u \,   \,  e^{i \tilde{\beta}} \right)   \right|   \, . \nonumber  
\end{align}
It can be noticed that for fixed values of the asymmetries $x_d$, $x_u$,
$V_{us}$ depends on the phase $\tilde{\delta}$ while $V_{cb}$ only on the phase $\tilde{\beta}$.
It is also worth noting that for $x_d=1$, the contribution of $s_{13}^d$ in $|V_{ub}|$ is negligible and the Fritzsch texture implies the prediction
$|V_{ub}/V_{cb}|\approx \sqrt{y_u / y_c } $.
Similar considerations can be inferred for the other off-diagonal elements,
 with the prediction $|V_{td}/V_{ts}|\approx \sqrt{y_d / y_s } $ for $x_d=1$.
As regards the complex part of $V_\text{CKM}$, 
we can consider the rephasing-invariant quantity
$J=-\text{Im}(V^*_{us}V^*_{cb}V_{ub}V_{cs}) $,
the Jarlskog invariant. In our scenario  we have
\begin{align}
 J= & -\sin\tilde{\delta} \, s_{12}^u s_{12}^d \, \left[  (s_{23}^d)^{2}\, c_{23}^d c_{12}^d 
- 2 \cos\tilde{\beta} \, s_{23}^d s_{23}^u + (s_{23}^u)^{2}  \right] +  \nonumber  \\
&\, +  \left( \sin\tilde{\delta} s_{12}^u s_{23}^d+ \sin\tilde{\beta} s_{12}^d s_{23}^u \right) s_{13}^d+\mathcal{O}(\epsilon_d^4) \, .
\label{Jamano}
\end{align}

In the following  we are going to test the viability of asymmetric Fritzsch textures 
given the present precision of experimental data on moduli and phases of the mixing elements.  
 The values of  the CKM matrix parameters and their uncertainties   
 are given in table \ref{values}.

\begin{table}[th]
\begin{tabular}{@{} lr  @{\hspace{4\tabcolsep}}  l r @{\hspace{4\tabcolsep}} lr @{}} 
Observable & Value & Observable & Value & Parameter & Global fit value \cite{Workman:2022ynf}  \\ \hline
$|V_{ud}|$ & $0.97373(31)$  & $|V_{cs}|$ & $0.975(6)$  &$\sin\theta_{12}$ & $0.22500\pm 0.00067$   \\
$|V_{us}|$ & $0.2243(8)$  &$|V_{cd}|$ & $0.221(4)$  & $\sin\theta_{23}$ & $0.04182^{+0.00085}_{-0.00074}$ \\
$|V_{ub}|$ & $0.00382(20)$  & $|V_{td}|$ & $0.0086(2)$ & $\sin\theta_{13}$ & $0.00369\pm 0.00011$ \\
$|V_{cb}|$ & $0.0408(14)$  & $|V_{ts}|$ & $0.0415(9)$ & $J$ & $(3.08^{+0.15}_{-0.13})\times 10^{-5}$ \\
$|V_{ub}/V_{cb}|$ & $0.084(7)$    &  $|V_{td}/V_{ts}|$ & $0.207(3)$  & $\delta$ & $1.144\pm 0.027$ \\
\hline
\end{tabular}
\caption{\label{values} Independent determinations of the CKM elements 
and result of the global fit of four CKM parameters 
with the  constraints implied by the unitarity of $V_\text{CKM}$, 
as reported by PDG \cite{Workman:2022ynf}. }
\label{values}  
\end{table}

\begin{figure}
\centering
\includegraphics[width=0.56\textwidth]{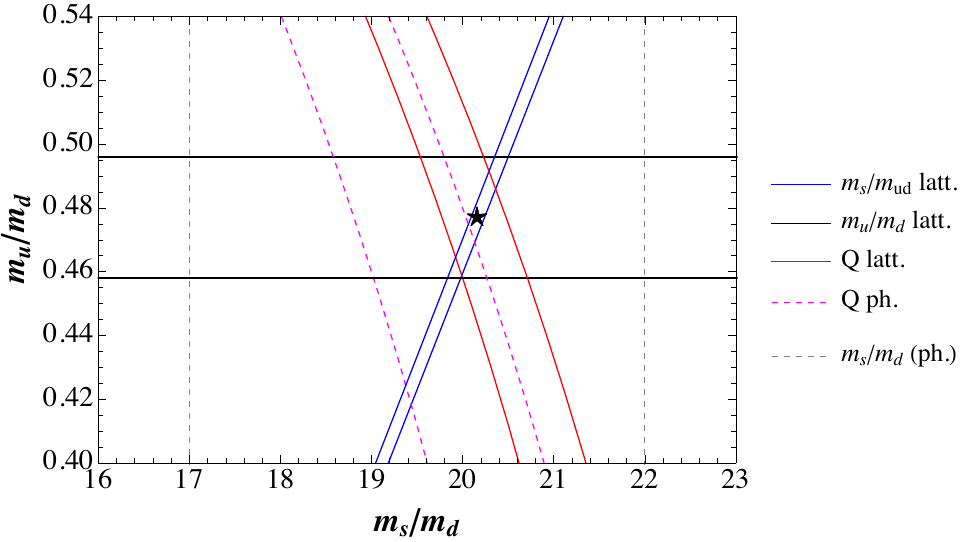}
\caption{ Light quarks mass ratios.  
Black solid lines show the average of the lattice determinations of $m_u/m_d$; 
blue lines represent the average of  the lattice determinations 
of the ratio $m_s/m_{ud}$, $m_{ud}=(m_u+m_d)/2$; red lines are obtained from the relation 
$Q^2=(m_s^2-m_{ud}^2)/(m_d^2-m_u^2)$, using lattice determinations of quark mass ratios. 
We also indicate the phenomenological result  
$Q= 22.1(7)$ \cite{Colangelo:2018jxw} (dashed magenta) and `old' limit 
$m_s/m_d = 17$--$22$ (dashed grey).
The black star represents the central value $(m_u/m_d,m_s/m_d )=(0.477, 20.17)$.
}
\label{mumd}
\end{figure}

The input values in our analysis are the ratios of the Yukawa eigenvalues. 
More specifically,  we want to reproduce  the CKM elements  as functions of the Yukawa 
couplings assuming that the matrices $Y_{u,d}$ acquire the Fritzsch form at some high energy scale
of new physics. Thus, we need to consider the evolution of Yukawa matrices  
according to the renormalization group (RG) equations. 
For that purpose, first we bring the running masses of all quarks 
to the scale $\mu = m_t \approx v_w$, using the QCD renormalization factors 
(see e.g. in \cite{Chetyrkin:2000yt}) and determine their ratios 
(i.e. the Yukawa constant ratios) at this scale. 
In doing so we obtain:   
\begin{align}\label{input-quarks}
& m_d/m_s = (20.17\pm 0.27)^{-1} \, , \qquad  m_s/m_b =  (53.94\pm 0.12)^{-1}    \nonumber \\
& m_u/m_c= (498 \pm 21)^{-1}   \, , \qquad \qquad  m_c/m_t = (272.3 \pm 2.6)^{-1} 
\end{align}
Here we used the PDG data on the top quark mass 
and the precision results of the lattice QCD computations  for the ratios 
$m_b/m_c= 4.579(9)$, $m_b/m_s = 53.94(12)$, 
$m_c/m_{s} = 11.768(34)$, $m_u/m_d  =  0.477(19)$ and   
$m_s/m_{ud} = 27.31(10) $,   where $m_{ud} = (m_u+m_d)/2$ \cite{Workman:2022ynf}.  
In particular, from the two latter results  we obtain the mass ratios of light quarks, 
$m_s/m_d=m_s/m_{ud} \, (m_u/m_d + 1)/2$  -- see fig.~\ref{mumd}.

The obtained pattern of the quark mass ratios \eqref{input-quarks}
can be compared to that of the lepton masses  known with much greater precision: 
$m_e/m_\mu = 206.76852^{-1}$ and $m_\mu/m_\tau= 16.817^{-1}$.  
Then we observe that the following combinations of down quark and lepton masses  
coincide with extremely good precision: 
\begin{align}\label{golden}
\frac{ \sqrt{m_em_\mu} }{m_\tau} = 241.819^{-1} \, , \qquad 
 \frac{ \sqrt{m_dm_s} }{m_b} = (242.2\pm 1.7) ^{-1}    
\end{align}
In the context of the Fritzsch-like textures \eqref{Yde} this relation implies 
$A_d/C_d = A_e/C_e$ which in turn stems from the $SU(5)$ symmetry 
in the context of $SU(5)\times SU(3)_H$ model in ref. \cite{Belfatto:2023qca}. 

 Now we are going to test the viability of the textures in eq. \eqref{fritzsch-mod} 
 when  $x_u=1$ but $x_d\neq 1$, i.e. when $\tilde{Y}_u$ has a 
`symmetric' form \eqref{Yu} while $\tilde{Y}_d$ is deformed as in \eqref{Yde}. 
In this case we have $9$ real parameters 
$A_{d,u}$, $B_{d,u}$, $C_{d,u}$, $\tilde{\beta}$, $\tilde{\delta}$ and  $x_{d}$, 
which have to match $10$ observables, 
the $6$ Yukawa eigenvalues and the $4$ independent parameters of the CKM matrix. 
 
In our numerical analysis the ratios of quark running masses at $\mu = m_t$  will be fixed 
as in eq. \eqref{mass-ratios}. However, we consider that 
 the `starting'  scale (i.e. the mass scale of heavy fermions \eqref{VL}) 
at which the Yukawa matrices acquire the Fritzsch form (\ref{fritzsch-mod}) can be much larger,  
say from $10^3$~GeV  to $10^{16}$~GeV.\footnote{
 The latter choice 
is natural in the context of the grand unified model $SU(5) \times SU(3)_H$,  
whereas the former  is interesting since 
the quark mixing with  the  vector-like quarks \eqref{VL} in the TeV mass range 
can be at the origin of
the recently observed Cabibbo angle anomalies \cite{Belfatto:2019swo} 
and can also induce various flavor-changing phenomena accessible in the SM precision tests 
\cite{Belfatto:2018cfo,Belfatto:2021jhf,Branco:2021vhs,Crivellin:2022rhw,Belfatto:2023tbv,Dcruz:2022rjg} 
(see also ref. \cite{Fischer:2021sqw} for a review). }
 For energy scales $\mu \gg m_t$, the quark mass ratios are no more given 
 by \eqref{mass-ratios} since their RG evolution will be influenced by additional contributions 
from  the top Yukawa constant $\sim 1$. 
However, we anticipate that these effects are small  and the obtained results are 
practically independent of the choice of the `starting' scale $\mu$ \cite{Belfatto:2023qca}.

\begin{figure}
\centering
\begin{subfigure}{0.48\textwidth}
\includegraphics[width=\textwidth]{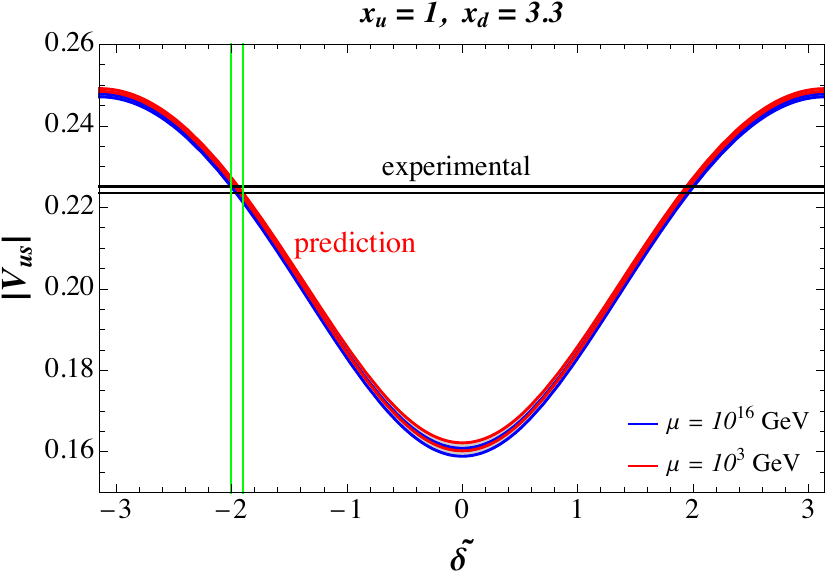}
\caption{\label{asymm1}}
 \end{subfigure}
 \hfill
\hspace{10pt}
\begin{subfigure}{0.48\textwidth}
\includegraphics[width=\textwidth]{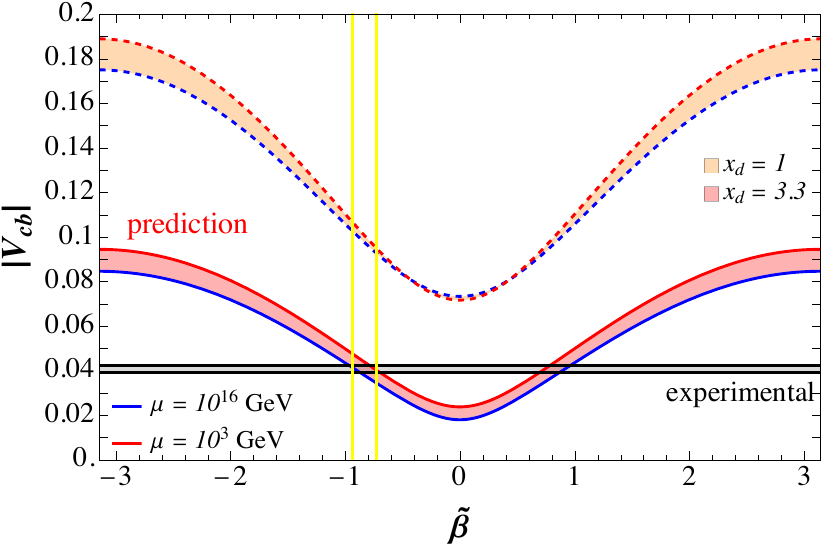}
\caption{\label{asymm2}}
 \end{subfigure}
 \hfill
 \\ \vspace{10pt}
 \begin{subfigure}{0.48\textwidth}
\includegraphics[width=\textwidth]{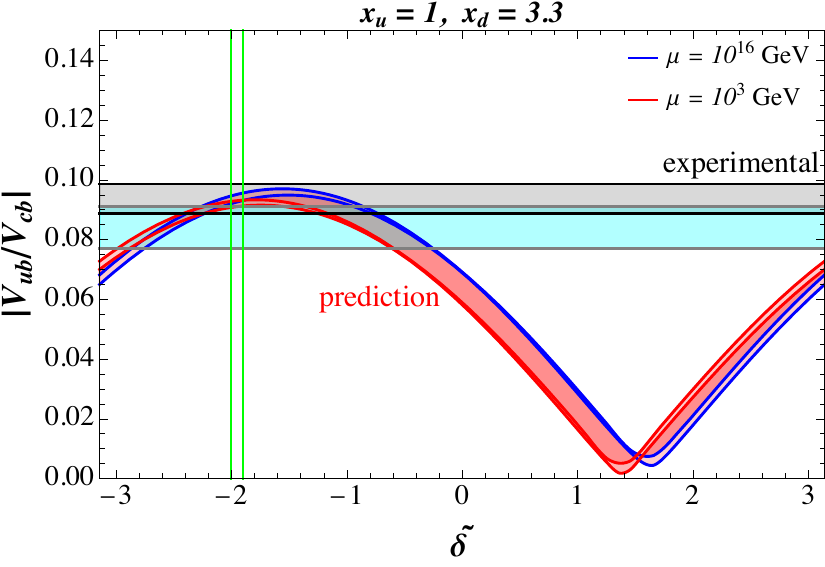}
\caption{\label{asymm3}}
 \end{subfigure}
\hfill
\begin{subfigure}{0.48\textwidth}
\includegraphics[width=\textwidth]{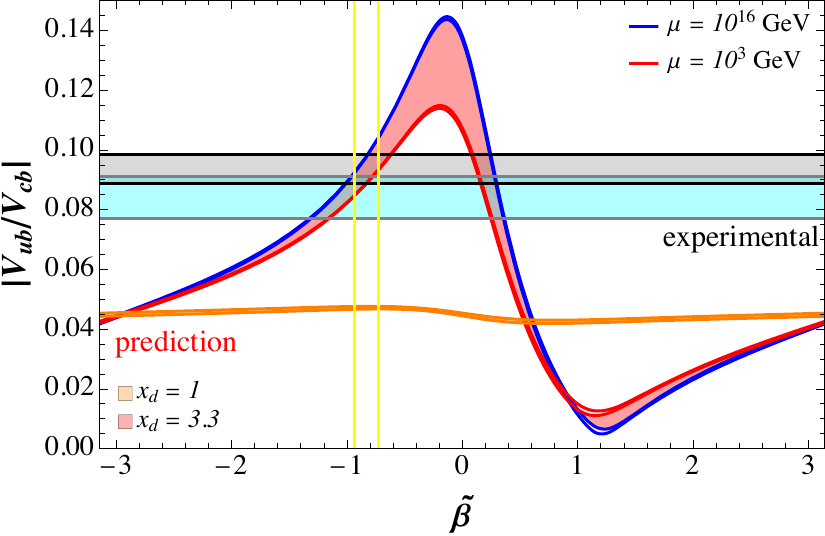}
\caption{\label{asymm4}}
 \end{subfigure}
   \hfill
 \\ \vspace{10pt}
\begin{subfigure}{0.48\textwidth}
\includegraphics[width=\textwidth]{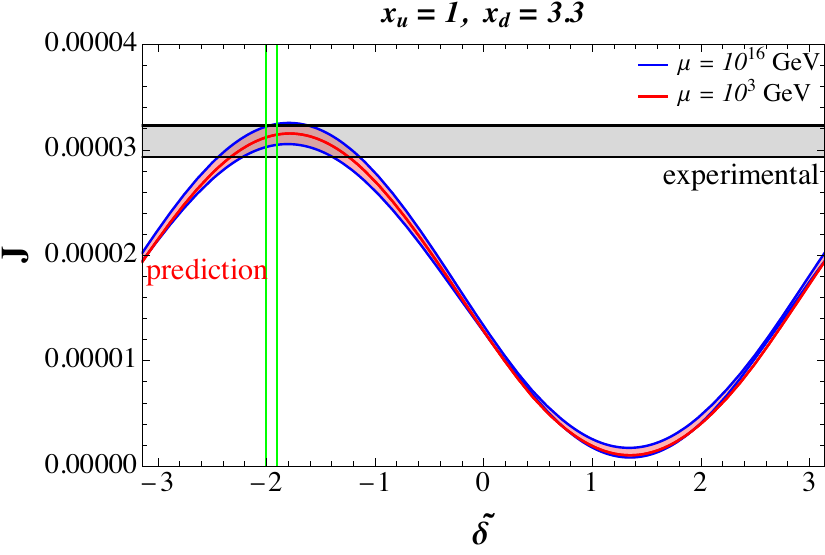}
\caption{\label{asymm5}}
 \end{subfigure}
 \hfill
\hspace{10pt}
\begin{subfigure}{0.48\textwidth}
\includegraphics[width=\textwidth]{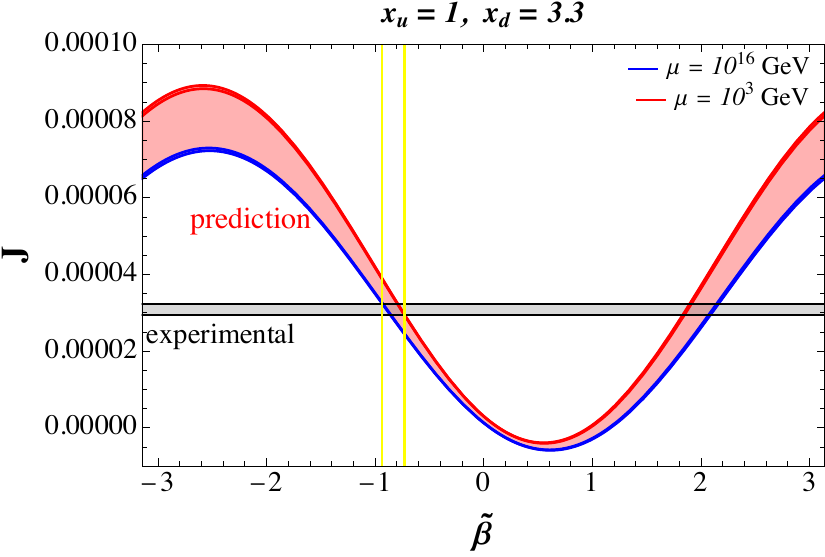}
\caption{\label{asymm6}}
 \end{subfigure} \caption{
Predicted observables (red bands) for Fritzsch-like texture with $x_d=3.3$, $x_u=1$ confronted with experimental 
values (horizontal bands) at $1\sigma$ confidence level.
The thickness of the bands of the prediction curves comprises the dependence on the starting scale $\mu = (10^3\div 10^{16})$ GeV.
In displaying the plots, the other variables are allowed to move inside the $1\sigma$ confidence region.
We also indicate the $1\sigma$ interval of the phases $\tilde{\delta}$ and $\tilde{\beta}$ (green and yellow bands respectively).
We also show the wrong predictions for $|V_{cb}|$ and $|V_{ub}/V_{cb}|$ with the symmetric Fritzsch textures (orange bands). 
}
\label{asymm}
\end{figure}

The complete analysis was done in ref. \cite{Belfatto:2023qca}. In  fig.~\ref{asymm} 
we show  its results for a choice $x_d=3.3$ which demonstrates 
that all CKM parameters can be obtained within their $1\sigma$ uncertainties 
for a proper choice of phases $\tilde{\delta}$ and $\tilde{\beta}$.

\begin{figure}
\centering
\begin{subfigure}{0.48\textwidth}
\includegraphics[width=\textwidth]{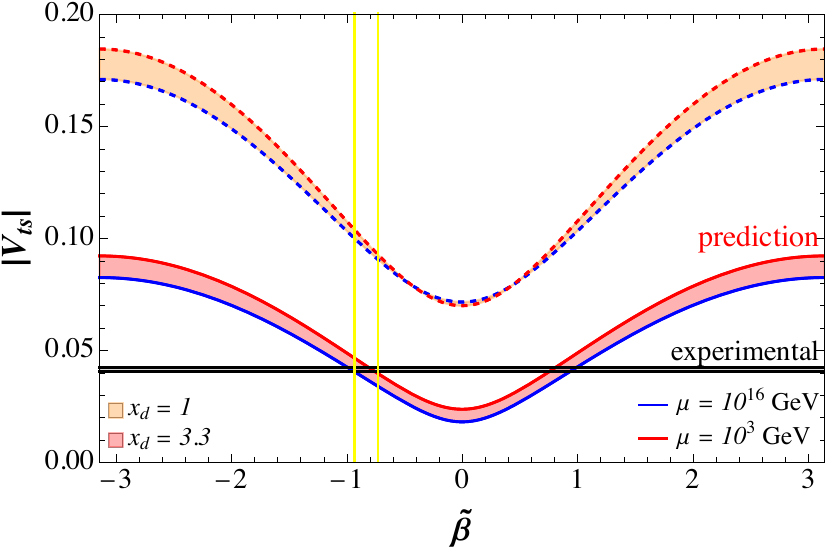}
\caption{\label{fig-vts}}
 \end{subfigure}
 \hfill
\hspace{10pt}
\begin{subfigure}{0.48\textwidth}
\includegraphics[width=\textwidth]{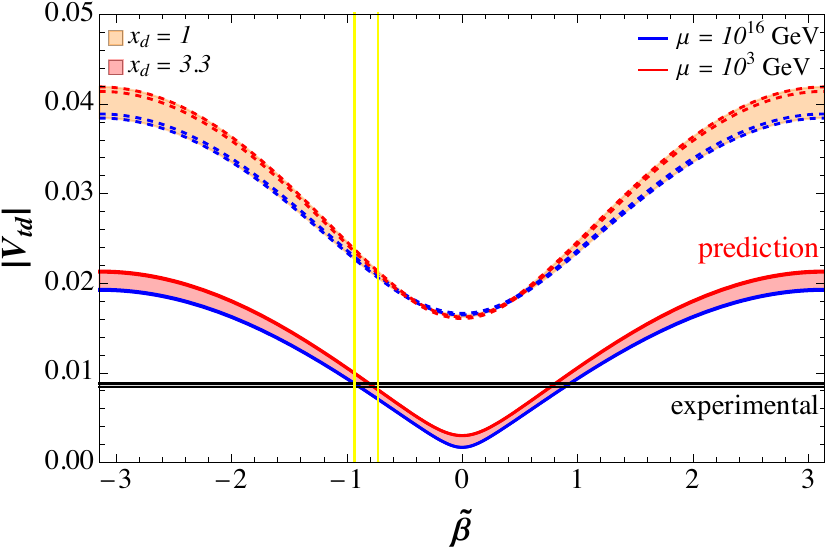}
\caption{\label{fig-vtd}}
 \end{subfigure}
 \caption{
Predictions of $V_{ts}$ and $V_{td}$ with $x_d=3.3$, $x_u=1$. 
We also show the wrong predictions of the symmetric Fritzsch texture (orange region).  
}
\label{fig-vtdvts}
\end{figure}

For having an insight of how it works, we can use the approximate relations in eqs. \eqref{vamanoeps} 
and \eqref{sd} (though in ref. \cite{Belfatto:2023qca} the compete analysis was done 
without any approximation). 
As it is apparent from eq. \eqref{vamanoeps},  the element $\vert V_{us} \vert = s_{12}$ 
(the Cabibbo angle) is  fixed by the phase $\tilde{\delta}$ and 
 has no dependence on $\tilde{\beta}$ in leading approximation. 
In other words, for the ratios $y_d/y_s$ and $y_u/y_c$ fixed as in \eqref{mass-ratios}, 
$\tilde{\delta}\approx \pm 2\pi/3 $ is fixed by the value of $s_{12}$ as shown in fig.~\ref{asymm}(a). 
Conversely, $\vert V_{cb}\vert =s_{23}$ depends only on $\tilde{\beta}$, 
and thus, for the given $x_d$, the latter phase is fixed as $\tilde{\beta}\approx \pm \pi/3 $
 by the value of $\vert V_{cb}\vert$ -- see fig.~\ref{asymm}(b).
 The sign of $\tilde{\delta}$ and $\tilde{\beta}$ is fixed by the other observables,
Namely, as one can see from the plots in fig.~\ref{asymm} the correct values of 
$|V_{ub}/V_{cb}|$ and $J$ can be obtained only for both phases being on the negative side  
(see the green and yellow vertical bands in fig.~\ref{asymm} respectively for $\tilde{\delta}$ and $\tilde{\beta}$). 
In fact, the effect of the asymmetry parameter $x_d>1$ is to decrease
the rotation angle $s_{23}^d$ while increasing $s_{13}^d$.
This causes the prediction of $|V_{cb}|$ to shift towards lower values. 
In addition, the large value of $s_{13}^d$ originates a dependence of $|V_{ub}|$ on both 
$\tilde{\delta}$ and $\tilde{\beta}$ so that for the `negative' choices of these parameters 
the value of $|V_{ub}/V_{cb}|$ intercepts the experimental determination.
 We also show in fig.~\ref{fig-vtdvts} how well the expectations on $V_{td}$ and $V_{ts}$ 
correspond to these phase parameters.

Concluding this discussion, we have shown that minimally deformed Fritzsch textures
with $x_u=1$ and $x_d\approx 3$ successfully predict  all observables of the CKM matrix. 
The choice of $x_u=1$ can be justified 
in the context of the grand unified picture $SU(5)\times SU(3)_H$,  
but $x_d$ remains a free parameter \cite{Belfatto:2023qca}.  
It is tempting to think that $x_d=3$  could also be originated 
from some more constrained scheme which is perhaps based on a larger GUT. 
Notice also that the results imply $\tilde{\delta} + \tilde{\beta }\approx -\pi $ 
which relation can be originated in the case of spontaneous CP-violation.


\section{Discussion: variations on the theme}
\label{conclusion}

The hierarchy between the fermion masses and their weak  mixing pattern 
remains a mystery. One can hope that  some new physics beyond the SM can shed a light  
on these problems.  However,  a complete theory which could to be coined as quantum flavordynamics 
does  not exist yet, and  one can discuss some pieces of the puzzle.  
The approach of zero-textures put forward by Fritzsch 
\cite{Fritzsch:1977vd,Fritzsch:1979zq} is oriented  towards the predictivity 
which in particular can relate the CKM mixing angles to the fermion mass ratios.   
Such predictive textures can be obtained within the models with the inter-family `horizontal' 
symmetries as we described in this paper,  but they do not address the origin of 
mass hierarchies: in fact, the hierarchies between the input parameters as in eq.
\eqref{ABC} can be related to the hierarchy 
of the horizontal symmetry breaking as in \eqref{VEV123}, but the latter is in fact 
introduced at hands. 
The predictive power of the textures can be enhanced by a clever use of 
the  grand unification theories as $SU(5)$ or $SO(10)$ unifying the fermions 
of one family. This can provide some successful relations between 
the quark and lepton masses as e.g. in eq. \eqref{AC}, but yet the explanation 
of the inter-family mass hierarchies remains beyond their reach. 

The fermion mass spectrum schematically given in fig.~\ref{plotmasse}
indeed looks very special. 
The only fermion with the mass of the order of electroweak scale $v_w=174$~GeV 
is the top quark, implying the Yukawa coupling $y_t \simeq 1$.  The other Yukawa 
couplings of the 3rd family are two orders of magnitude less,  $y_b, y_\tau \sim 10^{-2}$  
which indeed looks puzzling. Certainly, the coincidence of running couplings
$y_b(\mu)\simeq y_\tau(\mu)$ at the GUT scale $\mu \simeq 10^{16}$ GeV, 
the so called $b-\tau$ unification,  is a clear success of $SU(5)$, 
especially in its supersymmetric version, 
but the origin of $y_{b,\tau} \ll y_t$ still remains open. The question 
becomes more insistent in $SO(10)$ theory where $t,b$ and $\tau$ 
are unified in one multiplet, implying the Yukawa unification
$y_\tau = y_b = y_t$ at the GUT scale, and the only way to obtain the 
`vertical' splitting in 3rd family masses, $m_{b,\tau} \sim 10^{-2} m_t$,  
 is to introduce the large $\tan\beta$ at  hands. 
 The question is exacerbated by the fact that in the 2nd family the vertical 
 splitting is much less, $m_{s,\mu} \sim 10^{-1} m_c $,   
 Paradoxically, under the assumption of large $\tan\beta$, this would imply 
 the Yukawa $y_c$ an order of magnitude smaller than $y_{s,\mu}$,
namely $y_s \sim y_\mu \sim 10 y_c$. 
 This in turn implies $y_c \sim 10^{-2} y_{b,\tau}$, in obvious contrast 
 with the mass values of $c$-quark  (2nd family) and $b$-quark (3rd family)   
 which in fact are of the same order, $m_c \sim m_{b,\tau}$ (see fig.~\ref{plotmasse}).  
 Moreover, at the  normalization scale $\mu$ taken as the GUT scale,  
 the masses of the first family become quasi-degenerate: $m_d/m_u \simeq 2$ 
 while $m_e/m_u \simeq 1/2$, an additional difficulty for the large $\tan\beta$ 
 scenario. 
 
The overall pattern of fermion mass spectrum can be more naturally understood 
 in the inverse hierarchy approach \cite{Berezhiani:1983hm}
  which implies that the fermion mass hierarchies are inversely proportional 
  to the hierarchy of the inter-family symmetry breaking \eqref{VEV123}. 
 In fact,  fig.~\ref{plotmasse} shows  the following scaling laws  for inverse masses:
 \begin{align}\label{inverse} 
 m_u^{-1} : m_c^{-1} : m_t^{-1} \sim 1 : \epsilon_u : \epsilon_u^2, \qquad 
 m_d^{-1} : m_d^{-1} : m_b^{-1} \sim 1 : \epsilon_d : \epsilon_d^2 
 \end{align} 
 and similarly for leptons, with the near degeneracy    
 $m_d\simeq m_u \simeq m_e$ which indicates 
 that the Yukawa unification at the GUT scale may take place in the first   
  rather than in the third family, i.e. $y_d=y_u=y_e$ instead of  $y_b=y_t=y_\tau$.  
  This in turn suggests the following pattern for the inverse 
  Yukawa matrices \cite{Berezhiani:1995yk,Berezhiani:2005tp}:\footnote{
   For  quarks ($f=u,d$) such a pattern  was first obtained  
 in refs.  \cite{Berezhiani:1991ds,Berezhiani:1992pj,Berezhiani:1992bx} 
  in the context of left-right symmetric models.} 
 \begin{align}\label{Yinv} 
 Y_f^{-1} = y^{-1} (P_1 + \epsilon_f P_2 + \epsilon_f^2 P_3 ), \quad f=u,d,e 
 \end{align}  
 where $\epsilon_f = \epsilon_u,\epsilon_d,\epsilon_e$ are small parameters 
and  $P_{1,2,3}$ are symmetric rank-1 matrices with generically complex 
$O(1)$ elements.  
Without loosing generality, one can take   $P_1=(1,0,0)^T \bullet(1,0,0)$, 
$P_2 =  (a,b,0)^T \bullet(a,b,0)$ and $P_3 =  (x,y,z)^T \bullet(x,y,z)$. 
Therefore, the inverse Yukawa matrices have onion-like structures: 
\begin{align}\label{Y-onion} 
Y_{f}^{-1}  =  \frac{1}{y} \left(\begin{array}{ccc}
1 + a^2 \epsilon_f &ab \epsilon_f & xz \epsilon_f^2 \\ 
 ab \epsilon_f  & b^2 \epsilon_f  &  yz\epsilon_f ^2 \\ 
 xz \epsilon_f^2 & yz \epsilon_f^2 & z^2 \epsilon_f^2 \end{array}\right), \qquad f=u,d,e 
 \end{align}
Such structure of the Yukawa matrices  in the frame of 
$SO(10)$ symmetry was discussed in refs. \cite{Berezhiani:1995yk,Berezhiani:2005tp}.  
Three fermion families in representations $16^i$, 
acquire masses via their seesaw-like mixing with 
extra vector-like fermions  $16'_i + \overline{16}'_i$  
while the latter get masses via couplings 
 $16'_i \big[ \chi_1^{ij}  + (45/M) \chi_2^{ij} + (45/M)^2 \chi_3^{ij} \big] \overline{16}'_j $.
 In the above $45$ is a scalar in adjoint representation of $SO(10)$ 
 having VEV $M_G \sim 10^{16}$~GeV (the GUT scale),    
$M > M_G$ is some large (string?) scale, and 
 $\chi_{1,2,3}^{ij}\sim $ are flavon fields in symmetric (sextet) representations of $SU(3)_H$,  
having the  rank-1 VEV configurations in family space (as $\langle \chi_3 \rangle$ 
in \eqref{vevs}) disoriented by large angles \cite{Berezhiani:2005tp}.
In this way, after integrating out the heavy fermions, one obtains 
the quark and lepton Yukawa matrices of the form \eqref{Yinv}, with rank-1 projectors 
 originated from the flavon VEVs: $P_{n} = \langle \chi_n \rangle/M$, 
 and with small expansion parameters $\epsilon_f \sim M_G/M$.  
 The VEV $\langle 45 \rangle$ produces different Clebsch factors  
for different fermion species  so that generically
 $\epsilon_u$,  $\epsilon_d$ and $\epsilon_e$  in \eqref{Yinv}  have different values.    
 However, $SO(10)$ symmetry implies 
 a remarkable relation between the expansion parameters: 
 \begin{align}\label{eps-e} 
 \epsilon_e = - (\epsilon_d + 2\epsilon_u) \quad  \Longrightarrow \quad 
 \epsilon_e \approx - \epsilon_d 
 \quad \mathrm{since} \quad \epsilon_u \ll \epsilon_d 
 \end{align} 
Eigenvalues of  \eqref{Y-onion} reproduce the scaling pattern 
\eqref{inverse} implying that $\epsilon_u \ll \epsilon_d \ll 1$.  
However, the term $a^2 \epsilon_d$ in $Y_d$ cannot be neglected. 
Namely, for its eigenvalues we have $y_d \approx y/\vert 1+ a^2 \epsilon_d \vert $, 
$y_s \approx y \vert 1+ a^2 \epsilon_d \vert /\vert b^2\epsilon_d\vert$,  
$y_b \approx y/\vert z\epsilon_d\vert^2$ and similarly for leptons. 
However, eq.~\eqref{eps-e} implies that $y_d y_s \approx y_ey_\mu$ and 
$y_b \approx y_\tau$, and thus relation \eqref{golden}. 
As for up quarks, $a^2 \epsilon_u$ is negligible, and we have 
 $y_u\approx y$, $y_c \approx  y/\vert b^2\epsilon_d\vert$, $y_t \approx y/\vert z\epsilon_u\vert^2$. 
In addition, the CKM mixing is dominated by the diagonalization of $Y_d$. 
Then, taking into account that $ \epsilon_e \approx - \epsilon_d$, we get:
\begin{align}\label{good} 
\frac{y_u}{y_d} \approx \vert 1+ a^2 \epsilon_d \vert, \quad 
\frac{y_u}{y_e} \approx  \vert 1- a^2 \epsilon_d \vert, \quad 
s_{12} \approx \frac{ \vert ab \epsilon_d\vert}{\vert 1+ a^2 \epsilon_d \vert}
\approx \sqrt{\frac{y_d}{y_u} \vert a^2 \epsilon_d \vert }
\end{align} 
Thus, the last relation for for the Cabibbo angle implies that $\vert a^2 \epsilon_d \vert \simeq 1$ 
which allows to achieve mass splitting in first family, $m_u/m_d \approx 1/2$ 
and $m_u/m_e\approx 2$, by properly choosing the phase of $a^2\epsilon_d$. 
 As for other mixing angles, one  they naturally are $s_{23} \sim \epsilon_d$ and 
$s_{13} \sim \epsilon_d^2$ in a  fine correspondence with  \eqref{angles}.

In a completely different approach, the fermion mass pattern can be explained by 
 exceptionally clever choice of the underlying GUT,  
 even without introducing any type of flavor symmetry. 
 Namely,  supersymmetric grand unification based on $SU(6)$ \cite{Berezhiani:1989bd}
 solves the fundamental problems of grand hierarchy and doublet-triplet splitting 
in elegant way, since the Higgs doublets emerge as pseudo-Goldstone 
modes of the accidental global symmetry $SU(6) \times SU(6)$ in the Higgs superpotential 
\cite{Berezhiani:1989bd,Berezhiani:1995sb,Dvali:1996sr}. 
The mass parameters at low energies 
(soft supersymmetry breaking masses as well as supersymmetric $\mu$-term)  
are all originated from the SUSY breaking terms. 
The pseudo-Goldstone nature of the Higgs has important consequences: 
it can have the renormalizable coupling in the superpotential, with the Yukawa constant
$y_t \sim 1$ , only to one of the three up-type quarks that can be identified as the top quark 
\cite{Barbieri:1994kw,Berezhiani:1995dt}.  
For the rest of fermion species, the Yukawa couplings can emerge 
 only from the  higher order operators and thus are suppressed. 
 Interestingly, the structure of these operators also leads to 
 $y_c \sim y_{b,\tau} \ll y_t$, with a perfect match to small values of $\tan\beta$.  
 As for the mass hierarchy  between the 1st and 2nd families, it 
  cannot be explained without help of flavor symmetries, but some discrete 
  symmetries can be sufficient  \cite{Berezhiani:1995dt}.

\bigskip 

\section*{Postscriptum from Zurab Berezhiani}

I was a Ph.D. student when I encountered the works of Harald Fritzsch  
\cite{Fritzsch:1977vd,Fritzsch:1979zq}  which deeply influenced my further career. 
I started to think on the role of inter-family symmetry $SU(3)_H$  
for understanding the fermion masses and mixings. 
The first time I met Harald personally was in 1990 
in the last days of my one month visit to the Max-Planck Institute in Munich.  
During our conversations I discussed with him my papers 
\cite{Berezhiani:1983hm,Berezhiani:1985in} in which I succeeded to 
realize the Fritzsch mass textures in a theoretically appealing way  
by relating them to the pattern of $SU(3)_H$ symmetry breaking. 
As a consequence,  Harald invited me for a Humboldt fellowship to 
Ludwig-Maximilian University where I spent more than one year, 
and we never lost contacts after:  
he regularly invited me for seminars in Munich and for conferences 
that he organized worldwide, and many times he visited me in L'Aquila.     
Benedetta and me had discussions with Harald  
on the viable modifications of Fritzsch zero-texture during his last visit
which work was completed  later as ref.~\cite{Belfatto:2023qca}. 
 
Harald often asked me about Georgia and he was eager to visit Tbilisi. 
After COVID era, we decided to organize a conference  in Tbilisi on 
"Recent Advances in Fundamental Physics", planned for the fall 2022.  
But in August 2022 I received a sad news from Bigitte 
that Harald passed away.  
 
 I feel privileged to have had close scientific and personal ties with Harald Fritzsch.  
I remember him as a renowned physicist who authored many breakthrough works in 
 different areas of particle physics 
but also as a man with a pleasant and exceptionally humble personality. 
 
 \subsection*{Acknowledgements}

The work of Z.B. was partially supported by the research grant No.
2022E2J4RK "PANTHEON: Perspectives in Astroparticle and
Neutrino THEory with Old and New messengers" under the program PRIN 2022
funded by the Italian Ministero dell’Universit\`a e della Ricerca (MUR) and
by the European Union – Next Generation EU.
The work of B.B. was supported by the Deutsche Forschungsgemeinschaft (DFG, German Research Foundation) under grant 396021762 - TRR 257.


\end{document}